\documentclass[twocolumn,groupaddress,aps,prb,UTF8]{revtex4-2}
\usepackage{amsmath}
\usepackage{graphicx}
\usepackage{dcolumn}
\usepackage{color}
\usepackage{ulem}
\usepackage{bm}
\usepackage{braket}
\usepackage[colorlinks,citecolor=blue]{hyperref}
\usepackage{multirow}
\usepackage{makecell}
\usepackage{booktabs}
\usepackage{graphicx}
\usepackage{soul}

\begin{document}

\title{Non-Adiabatic Effect in Topological and Interacting Charge Pumping}
\author{Fan Yang}
\affiliation{Institute for Advanced Study, Tsinghua University, Beijing,100084, China}
\author{Xingyu Li}
\affiliation{Institute for Advanced Study, Tsinghua University, Beijing,100084, China}
\author{Hui Zhai}
\email{huizhai.physics@gmail.com}
\affiliation{Institute for Advanced Study, Tsinghua University, Beijing,100084, China}
\date{\today}

\begin{abstract}

Topological charge pumping occurs in the adiabatic limit, and the non-adiabatic effect due to finite driving velocity reduces the pumping efficiency and leads to deviation from quantized charge pumping. In this work, we discuss the relation between this deviation from quantized charge pumping and the entanglement generation after a pumping circle. In this simplest setting, we show that purity $\mathcal{P}$ of the half system reduced density matrix equals to $\mathcal{R}$ defined as $(1-\kappa)^2+\kappa^2$, where $\kappa$ denotes the pumping efficiency. In generic situations, we argue $\mathcal{P}<\mathcal{R}$ and the pumping efficiency can provide an upper bound for purity and, therefore, a lower bound for generated entanglement. To support this conjecture, we propose a solvable pumping scheme in the Rice--Mele--Hubbard model, which can be represented as a brick-wall type quantum circuit model. With this pumping scheme, numerical calculation of charge pumping only needs to include at most six sites, and therefore, the interaction and the finite temperature effects can be both included reliably in the exact diagonalization calculation. The numerical results using the solvable pumping circle identify two regimes where the pumping efficiency is sensitive to driving velocity and support the conjecture $\mathcal{P}<\mathcal{R}$ when both interaction and finite temperature effects are present.

\end{abstract}

\maketitle

\section{Introduction}
Quantized transport is one of the most direct manifestations of nontrivial topology in quantum systems. It includes two different approaches. One is the quantized linear response to an external constant voltage field, including the quantized conductance in one-dimensional ballistic metals \cite{Landauer57,Kane22} and various kinds of quantum Hall effects \cite{vonKlitzing80,Tsui82,Haldane88,QAHexp,QAH,AQH,Kane05,Bernevig06,QSHexp,QSH}. Such measurements have been extensively used in condensed matter systems to probe topological phases. Generally speaking, this scheme requires a small voltage field such that the system remains in the linear response regime. The other is quantized charge transport in a dynamically driven process with periodically driving parameters \cite{Laughlin81,Thouless83,Niu84,Niu90,Xiao,Citro23}. The canonical example is the Thouless pump \cite{Thouless83}. Such measurements are easier to be realized in ultracold atomic gases \cite{Lohse16,Nakajima16,Lu16,Tangpanitanon16,Walter22,Dreon22}. Generally speaking, this scheme requires a slow driving velocity such that the entire dynamic process remains within the adiabatic limit. In both cases, a small external field or slow dynamics is necessary for manifesting topology. Deviating from linear response or adiabatic limit inevitably causes deviation from topological quantization. 

However, in practice, the driving speed is always finite in experiments, and consequently, the non-adiabatic effect results in a deviation from quantization. This deviation has been observed in a recent cold atom experiment \cite{Walter22}. In this experiment, there also presents tunable interactions between particles. The sensitivity to finite driving speed depends on interaction as well as temperature. Previously, in the case of crossing a symmetry-breaking phase transition, a finite ramping rate leads to the generation of topological defects whose density reveals the critical exponents, known as the celebrated Kibble--Zurek mechanism \cite{Kibble76,Kibble80,Zurek,KZ-review}. Response of a physical observable to a finite driving rate has also been used to probe intrinsic many-body correlations \cite{Hu}. Nevertheless, theoretical investigations of the non-adiabatic effect in topological charge pumping are still limited \cite{Niu90,Shih94,Privitera,Grabarits}. Here we would like to understand whether the non-adiabatic effect can characterize the intrinsic properties of this physical system after pumping, such as entanglement generation \cite{Gawatz22}, and whether there exists a reliable way to compute the non-adiabatic effect in an interacting system. 

\begin{figure}
\centering
\includegraphics[width=\columnwidth]{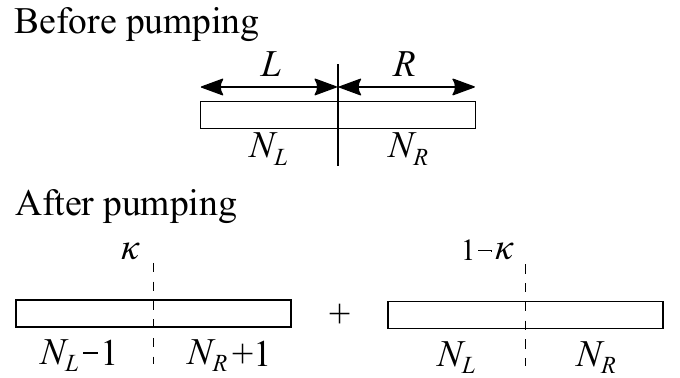}
\caption{Schematic of the relation between entanglement/density matrix purity generated by pumping and the pumping efficiency $\kappa$.}
\label{purity}
\end{figure}

\section{Relation between Purity and Pumping Efficiency} To begin with, let us first discuss an intuition connecting entanglement entropy generated by pumping and the non-adiabatic effect. As shown in Fig. \ref{purity}, we consider a one-dimensional chain and divide it into the left and the right halves, whose particle numbers are denoted by $N_\text{L}$ and $N_\text{R}$, respectively. In the adiabatic limit, a quantized charge is pumped from the left to the right, and the particle numbers become $N_\text{L}-1$ and $N_\text{R}+1$, respectively. However, in the non-adiabatic case, let us denote $\kappa$ as the pumping efficiency, that is to say, there are certain possibilities, approximated by $1-\kappa$, that the charge is not pumped and the particle numbers remain $N_\text{L}$ and $N_\text{R}$ after a pumping circle, as shown in Fig. \ref{purity}. 

If we take a partial trace over the right system and obtain the reduced density matrix for the left system $\rho_\text{L}$, we can introduce the second R\'enyi entropy $S_2=-\ln\text{Tr}\rho^2_\text{L}$ to describe the entanglement between two sides. Here $\text{Tr}\rho^2_\text{L}$ is also called the purity, denoted by $\mathcal{P}$. For noninteracting type Gaussian states, there exists a formula relating purity  with particle numbers. The formula reads \cite{Tam22}
\begin{equation}\label{purityEq}
\text{Tr}\rho^2_\text{L}=\prod\limits_{p=-1/2}^{1/2}\text{Tr}(\rho_\text{L} e^{i\pi p\hat{N}_\text{L}}).
\end{equation} 
Using this formula, we obtain
\begin{align}
\mathcal{P}=\text{Tr}\rho^2_\text{L}\approx&[(1-\kappa)e^{i\frac{\pi}{2} N_L}+\kappa e^{i\frac{\pi}{2}(N_L-1)}]\nonumber\\
&\times [(1-\kappa) e^{-i\frac{\pi}{2} N_L}+ \kappa e^{-i\frac{\pi}{2}(N_L-1)}]\nonumber\\
=&(1-\kappa)^2+\kappa^2. \label{relation}
\end{align}
This establishes the relation between purity and pumping efficiency. For quantized charge pumping, $\kappa$ is either unity or zero, leading to $\mathcal{P}=1$ and no generation of entanglement entropy. When the non-adiabatic effect occurs, $\kappa$ can take any value between zero and unity, reducing purity and generating entanglement. 

To arrive at this relation, we have assumed that the initial state has fixed particle numbers for both left and right sides. Strictly speaking, this relation also does not hold when interaction and finite temperature effects are taken into account. 
However, we will show that $\mathcal{P}$ and $\mathcal{R}$ exhibit the same trend, where $\mathcal{R}$ denotes $(1-\kappa)^2+\kappa^2$ as the right-hand side of Eq. \ref{relation}. 

Let us denote the probability of transporting $n$ charges as $P_n$. The discussion above only considers $P_0$ and $P_1$. Suppose we would like to consider other possibilities ignored above, and the next order contribution should be two particles being pumped or the back-flow of one particle from right to the left, which are denoted by $P_2$ and $P_{-1}$, respectively. Including all four possibilities and following Eq. \ref{purityEq}, we have 
\begin{equation}
\mathcal P= (P_{1}-P_{-1})^2+(P_0-P_2)^2.
\end{equation}
On the other hand, we have the pumping efficiency as  
\begin{equation}
\kappa=\sum_n nP_n=2P_2+P_1-P_1.
\end{equation}
and therefore $\mathcal{R}$ is given by
\begin{equation}
\mathcal R=(2P_2+P_1-P_{-1})^2+(P_0-P_2+2P_{-1})^2.
\end{equation}
It is easy to verify that
\begin{equation}
\mathcal{R-P}=4[P_{-1}P_0+P_1P_2+(P_{-1}-P_2)^2]>0.
\end{equation}
Hence, we further conjecture that $\mathcal{P}<\mathcal{R}$ as long as non-adiabatic effect is still perturbative. That is to say, the pumping efficiency provides an upper bound for purity and a lower bound for the entanglement generated during pumping.  

\section{Rice--Mele--Hubbard Model and A Solvable Pumping Cycle} 
The Rice--Mele model is the most widely used model for studying charge pumping \cite{Rice82}. By adding interactions, we consider the Rice--Mele--Hubbard model at half filling \cite{Nakagawa18,Bertok22}
\begin{equation}
\begin{split}
H=&-\sum_{i,\sigma}(J+(-1)^{i+1}\delta)(c_{i,\sigma}^\dagger c_{i+1,\sigma}+h.c)\\
&+\Delta\sum_{i,\sigma}(-1)^{i+1}n_{i\sigma}+U\sum_i n_{i,\uparrow}n_{i,\downarrow},
\end{split}
\end{equation}
where $n_{i,\sigma}=c_{i,\sigma}^\dagger c_{i,\sigma}$, $c_{i,\sigma}$ and $c^\dagger_{i,\sigma}$ are fermionic annihilation and creation operators. $J\pm \delta$ is the hopping amplitude between neighboring sites, alternating between two neighboring bonds. The odd and even sites form two sets of sublattices, and $\Delta$ denotes the energy detuning between them.  $U$ is the on-site interaction strength between different spins. In the pumping process, $J$, $\delta$, and $\Delta$ are functions of time $t$. We set $\hbar=k_B=1$, with $\hbar$ and $k_B$ being the reduced Planck constant and the Boltzmann constant, respectively. 

\begin{figure}
\centering
\includegraphics[width=0.8\columnwidth]{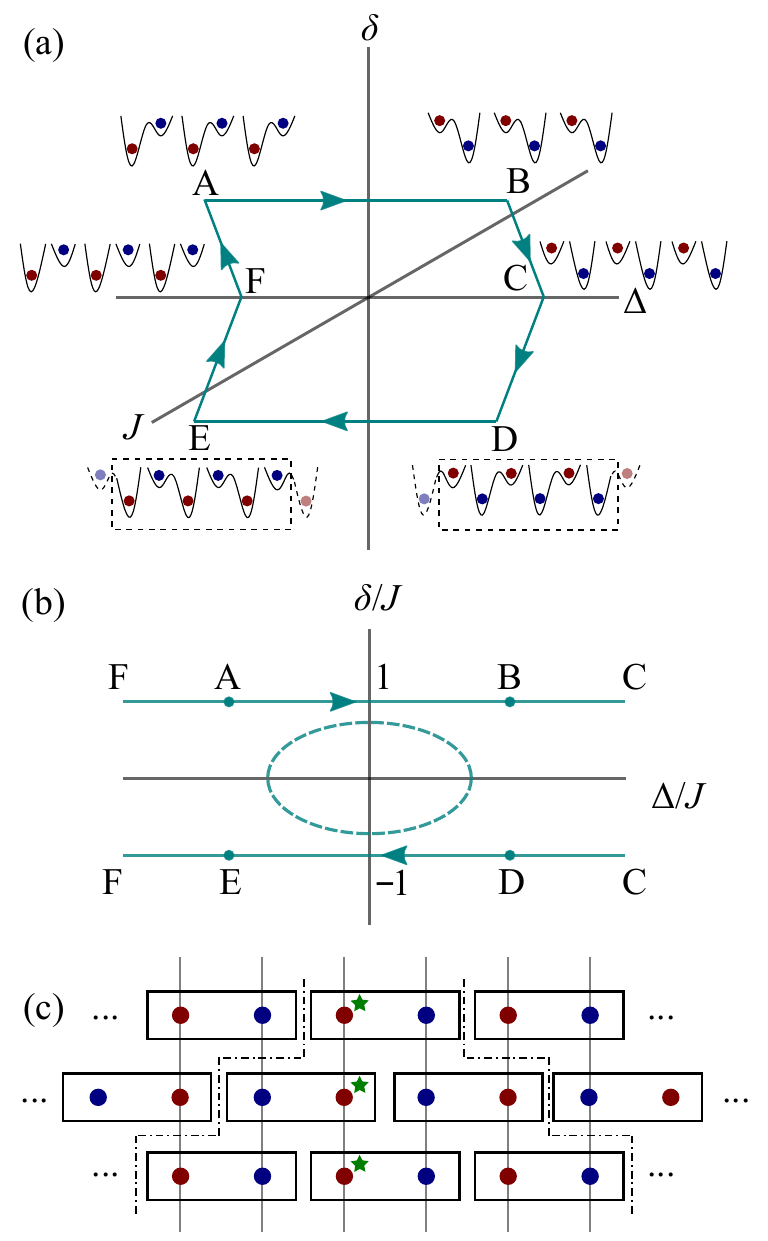}
\caption{Illustration of the solvable pumping cycle. (a) The trajectory of the pumping cycle in the $\delta-\Delta-J$ parameter space. Here $B\rightarrow C$ and $F\rightarrow A$ lines obey $\delta=J$ and $C\rightarrow D$ and $E\rightarrow F$ lines obey $\delta=-J$. The lattice potentials are illustrated at the starting points of each step, where the hopping between two adjacent sites is nonzero (zero) if they are connected (disconnected). (b) The same trajectory as (a), but shown in the $\delta/J-\Delta/J$ parameter space. For comparison, the dashed line shows the conventional topological pumping trajectory, which can be continuously stretched to our trajectory. (c) This pump protocol is represented by a three-layer brick-wall quantum circuit. The first layer represents $A\rightarrow C$, the second layer represents $C\rightarrow F$, and the third layer represents $F\rightarrow A$. The charge transport during one pumping cycle through a given lattice site, say, that marked by a star, can be computed exactly from only six sites within the information light cone denoted by the dash-dotted line.}
\label{illustration}
\end{figure}

In the adiabatic limit, for the non-interacting case, a quantized charge pumping can be observed when the trajectory of parameters $\delta$ and $\Delta$ encloses the gap closing point $\delta=\Delta=0$ in a pumping circle. This quantized charge pumping is stable against interaction until the repulsive Hubbard interaction $U$ exceeds $2|\Delta_0|$. Then, the charge pumping breaks down and the pumped charge is always zero no matter what parameter trajectory is chosen  \cite{Nakagawa18,Bertok22,Walter22}.

Charge pumping with finite driving velocity is a challenging dynamical problem when both interaction and finite temperature effects need to be considered. To this end, we will introduce a pumping scheme allowing us to compute the dynamics in a numerically exact way. This pumping scheme is illustrated in Fig. \ref{illustration}(a) and contains the following steps:

1. $A\rightarrow B$: We set $J_0=J(t=0)=1$. All energies are measured in units of $J_0$, and the pump velocity $v$ is measured in units of $J_0/\hbar$. We start with $\delta_0=\delta(t=0)=1$ under which the system forms disconnected double wells. Initially $\Delta_0=\Delta(t=0)<0$ and we linearly ramp $\Delta$ as $\Delta=\Delta_0(1-2vt)$ for a time duration $t_1=1/v$ to $\Delta=-\Delta_0>0$, with $J$ and $\delta$ fixed during the process. 

2. $B\rightarrow C\rightarrow D$: We fix $\Delta=-\Delta_0$ and linearly ramp $J$ and $\delta$ to zero as $J=J_0[1-v(t-t_1)]$ and $\delta=\delta_0[1-v(t-t_1)]$. In this process we keep the relation $J=\delta$ such that the system remains as disconnected double wells. When $t=t_2=t_1+1/v$, we reach the point $J=\delta=0$ and all sites are disconnected. After that, we ramp up $J$ as $J=J_0 v(t-t_2)$ and $\delta$ as $\delta=-\delta_0 v(t-t_2)$ until we reach $J=J_0$ and $\delta=-\delta_0$ at $t_3=t_2+1/v$. In this process, we always have $J=-\delta$, and the system is also disconnected double wells. However, these disconnected double wells are different from those in $A\rightarrow B\rightarrow C$. Suppose site-$i$ is connected to site-$(i+1)$ during $A\rightarrow B\rightarrow C$, site-$i$ is connected to site-$(i-1)$ and site-$(i+1)$ is connected to site-$(i+2)$ during $C\rightarrow D$. 

3. $D\rightarrow E$: We ramp $\Delta$ from $-\Delta_0$ to $\Delta_0$ with the same velocity $v$ and by keeping $J$ and $\delta$ fixed. 

4. $E\rightarrow F\rightarrow A$: This process reverts $B\rightarrow C\rightarrow D$ with $\Delta$ fixed. After this step, the parameters return to their initial values and a pumping circle is closed. 

Intuitively, charge transports from site-$i$ to site-$(i+1)$ from $A$ to $B$. Then, site-$i$ and $i+1$ are disconnected and site-$(i+1)$ is connected to site-$(i+2)$, which allows subsequently charge transport from site-$(i+1)$ to site-$(i+2)$ during $D$ to $E$. In such a way, unidirectional charge transport is realized. In this setting, it is straightforward to see that the quantized charge is stable for $U< 2|\Delta_0|$.

We remark on the difference between our pumping circle and the conventional ones. Normally, one fixes $J$ and tunes $\delta/J$ and $\Delta/J$ to form a closed loop during a pumping circle. In the non-interacting case, the pumping is topological or not depending on whether the trajectory loop in the $\delta/J-\Delta/J$ diagram encloses zero. As we deform the trajectory, topological pumping remains topological as long as no gap closing occurs in the loop. Here, although we change $J$ during the pumping, we can still plot the parameter trajectory in the $\delta/J-\Delta/J$ diagram, as shown in Fig. \ref{illustration}(b). Especially from $B$ to $C$, it corresponds to fixing $\delta/J=1$ and tuning $\Delta/J$ to positive infinity. Then, from $C$ to $D$, $\delta/J$ is fixed to $-1$, and $\Delta/J$ is tuned from infinity to finite. In other words, in the $\delta/J-\Delta/J$ diagram, we stretch the trajectory loop to $\Delta/J=\pm \infty$ and the parameter trajectory becomes two parallel straight lines. However, during this stretching process, no gap closing happens. Hence, our pumping circle is still topologically protected.   

A key feature of this designed pumping cycle is that the system always constitutes of disconnected double wells at any given time. Therefore, it can be represented by a three-layer brick-wall quantum circuit, as shown in Fig. \ref{illustration}(c). The first layer represents steps $A$ to $C$, the second layer $C$ to $F$, and the last layer $F$ to $A$. During one pump cycle, the pumped charge equals the net current passing through any given site. As shown in Fig. \ref{illustration}(c), a given site, marked by a star in Fig. \ref{illustration}(c), is only influenced by at most six sites within the light cone, as indicated by the dash-dotted line in Fig. \ref{illustration}(c). Therefore, we only need to compute the six sites by exact diagonalization, and we can obtain an exact result of charge pumping, including interaction, finite driving rate, and finite temperature effects.  

In practice, we compute the current during each step from the change of particle number $\Delta n$ at a given site $i$ labeled by the star in Fig. \ref{illustration}(c). Since the system always consists of disconnected double wells, the current and $\Delta n$ are equal (opposite) when site $i$ is connected to site $i-1$ (site $i+1$). 
When computing the pumped charge, we do not need to assume boundary conditions. As shown in the brick-wall circuit, at any given time the sites outside the light-cone denoted by the dash-dotted lines are always disconnected from site $i$ and do not contribute to $\Delta n$. Therefore, diagonalizing the six-site problem is equivalent to solving an infinite chain.

\begin{figure}
\centering
\includegraphics[width=0.9\columnwidth]{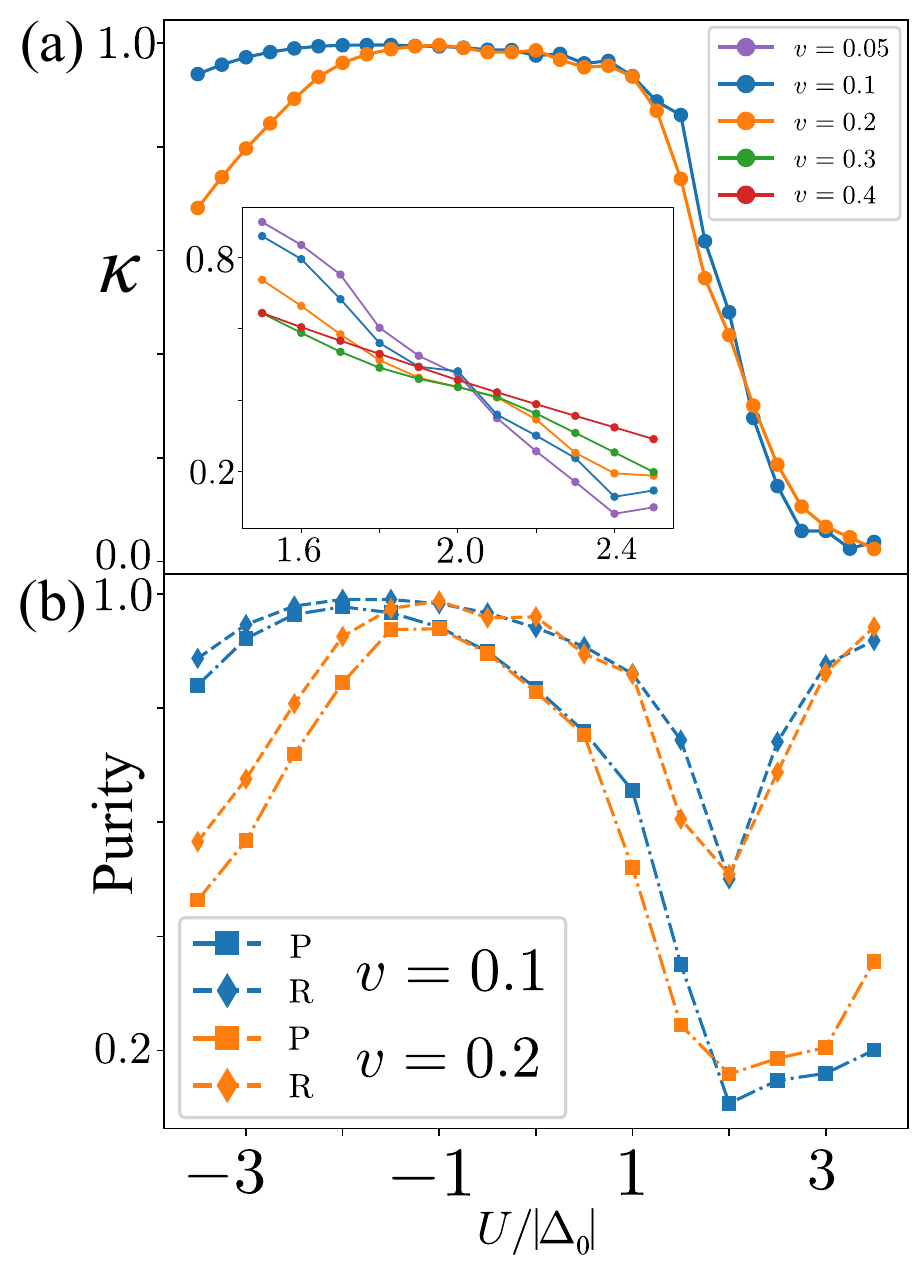}
\caption{Numerical results of charge pumping for the Rice--Mele--Hubbard model using the solvable pumping circle shown in Fig. \ref{illustration}. Here we fix $J_0=\delta_0=1$, $\Delta_0=-2$ and temperature $T=1$. (a) The pump efficiency $\kappa$ as a function of $U/|\Delta_0|$ for various driving speed $v$. The inset shows a zoom-in plot around $U=2|\Delta_0|$ where the pumping efficiency with different driving velocities nearly cross at a single point. (b) Comparison between purity $\mathcal{P}$ and $\mathcal{R}=(1-\kappa)^2+\kappa^2$ for two different driving velocities.}
\label{RMH}
\end{figure}

\section{Results} With the pumping circle introduced above, typical numerical results on the charge pumping on the Rice--Mele--Hubbard model is shown in Fig. \ref{RMH} \cite{source}. We have fixed $J_0=\delta_0=1$, $\Delta_0=-2$, $-7\leq U\leq 7$, and $T=1$. We have also varied temperature range and find that the results are not sensitive to temperature for the parameters we considered. 

First, we plot the pump efficiency in Fig. \ref{RMH}(a) for different driving velocities $v$. We find two regimes where the pumping efficiency strongly depends on velocity. One regime is $U<-2|\Delta_0|$. In this regime, attractive interaction binds two fermions with different spins to form a tightly bound pair, and the hopping of pairs scales as $\sim 1/|U|$. Hence, when the interaction is attractive and large, the kinetic energy of fermion pairs becomes small. Therefore, a small driving velocity can yield a large deviation. Another regime is $U\sim 2|\Delta_0|$. As shown in the inset of Fig. \ref{RMH}(a), when $U\lesssim 2|\Delta_0|$, the pumping efficiency $\kappa$ decreases as $v$ increases; while when $U\gtrsim 2|\Delta_0|$, $\kappa$ increases as $v$ decreases. Hence, when we plot $\kappa$ as a function of $U$ for different $v$, these curves cross nearly at the same point $U=2|\Delta_0|$. This point marks the transition from topological to non-topological charge pumping in the adiabatic limit, which is rooted in the ground state phase transition from a band insulator to a Mott insulator at $U=2|\Delta_0|$. In other words, we show that even with the experimental data measured at finite velocity, one is able to accurately locate the transition point of quantized charge pumping in the adiabatic limit. 

Here we should also note the difference between pumping efficiency and the deviation from quantization. When $U\lesssim 2|\Delta_0|$, topological pumping is quantized to unity and $\kappa$ is always greater than $0.5$. The deviation is $1-\kappa$. When $U\gtrsim 2|\Delta_0|$, pumping is quantized to zero in the adiabatic limit, and the deviation is $\kappa$, which is always smaller than $0.5$. The largest deviation from quantization is $0.5$ occurring at $U= 2|\Delta_0|$, where $\mathcal{R}$ reaches its minimum as $0.5$.

Finally, in Fig. \ref{RMH}(b), we support our conjecture $\mathcal{P}<\mathcal{R}$ by Fig. \ref{RMH}(b).  When computing purity, we assume open boundary condition. We compare $\mathcal{P}$ and $\mathcal{R}$ for two different driving velocities $v=0.1$ and $v=0.2$. We have calculated the results for a number of different temperature and velocities. All results are similar and not repeatedly shown here. We can see that $\mathcal{R}$ also decreases significantly as $v$ increases in the regime $U<-2|\Delta_0|$. Meanwhile, $\mathcal{R}$ exhibits a minimum around $U\approx 2|\Delta_0|$ where the deviation from quantization is the largest. This shows that $\mathcal{P}$ and $\mathcal{R}$ exhibit the same trend. Meanwhile, we also see that $\mathcal{P}$ is always smaller than $\mathcal{R}$. 

\section{Conclusion} In summary, we propose a pumping scheme that allows us to compute charge pumping for interacting cases in a numerically exact way. This scheme is also equivalent to a quantum circuit model, potentially bringing out the connection between charge pumping and digital quantum computing. We discuss the connection between deviation from quantization and generation of entanglement, both rooted in non-adiabatic dynamics. We also note that the relation between charge transport and entanglement has also been noticed in quantum point contact \cite{Klich09} and recently in quantized non-linear conductance \cite{Tam22,Kane22,Yang22,Zhang23}. All these phenomena might point to a unified picture behind them. 

\begin{acknowledgments} We thank Pengfei Zhang, Haifeng Tang, Chengshu Li, Minggen He, Wei Zheng, Chang Liu, and Tianshu Deng for helpful discussions. The project is supported by Beijing Outstanding Young Scholar Program, Innovation Program for Quantum Science and Technology 2021ZD0302005, NSFC Grant No.~11734010 and the XPLORER Prize. F.Y. is also supported by Chinese International Postdoctoral Exchange Fellowship Program (Talent-introduction Program) and Shuimu Tsinghua Scholar Program at Tsinghua University.
\end{acknowledgments}

\bibliographystyle{unsrt}

\end{document}